\documentclass[aps, prl, twocolumn, superscriptaddress, showpacs]{revtex4}
\usepackage{graphicx}

\begin{document}

\title{An Enhanced Nonlinear Critical Gradient  for Electron Turbulent Transport due to Reversed Magnetic Shear}
\author{J. L. Peterson}
\email{jpeterso@pppl.gov}
\affiliation{Princeton Plasma Physics Laboratory, Princeton University, Princeton, New Jersey 08543, USA}
\author{G. W. Hammett}
\affiliation{Princeton Plasma Physics Laboratory, Princeton University, Princeton, New Jersey 08543, USA}
\author{D. R. Mikkelsen}
\affiliation{Princeton Plasma Physics Laboratory, Princeton University, Princeton, New Jersey 08543, USA}
\author{H. Y. Yuh}
\affiliation{Nova Photonics Inc., Princeton, New Jersey 08540, USA}
\author{J. Candy}
\affiliation{General Atomics, San Diego, California 92186-2608, USA}
\author{W. Guttenfelder}
\affiliation{Princeton Plasma Physics Laboratory, Princeton University, Princeton, New Jersey 08543, USA}
\author{S. M. Kaye}
\affiliation{Princeton Plasma Physics Laboratory, Princeton University, Princeton, New Jersey 08543, USA}
\author{B. LeBlanc}
\affiliation{Princeton Plasma Physics Laboratory, Princeton University, Princeton, New Jersey 08543, USA}
\date{\today}

\begin{abstract}
The first nonlinear gyrokinetic simulations of electron internal transport barriers (e-ITBs) in the National Spherical Torus Experiment 
show that reversed magnetic shear can suppress thermal transport by increasing the nonlinear critical gradient for 
electron-temperature-gradient-driven turbulence to three times its linear critical value. An interesting feature of this turbulence is 
nonlinearly driven off-midplane radial streamers. This work reinforces the experimental observation that magnetic shear is likely an 
effective way of triggering and sustaining e-ITBs in magnetic fusion devices.
\end{abstract}


\maketitle

We aim to investigate, via simulation, recent experimental evidence on the National Spherical Torus Experiment (NSTX) 
\cite{ono_exploration_2000} of reduced electron thermal transport during reversed magnetic shear discharges 
\cite{yuh_suppression_2011}.

Understanding and mitigating thermal transport is paramount to the development of efficient and economical magnetic fusion reactors. 
In NSTX,  turbulent electron thermal flux can be very large, often serving as the experiment's dominant loss mechanism 
\cite{kaye_confinement_2007, kaye_scaling_2007}. 
Yet, by 
reversing the magnetic shear (as defined by $\hat{s}<0$, with $\hat{s}\doteq(r/q)\partial q/\partial r$, $r$ the machine's minor radial 
coordinate and $q$ its safety factor), NSTX can reach electron temperature gradients that far exceed the limits imposed by the onset of 
unstable electron-temperature-gradient-driven (ETG) turbulence \cite{yuh_internal_2009}. The present work investigates this 
phenomenon through the first nonlinear gyrokinetic simulations of strongly reversed shear NSTX plasmas. We find that reversed 
magnetic shear, even in the absence of $\mathbf{E}\times\mathbf{B}$ shear, suppresses electron turbulence by greatly enhancing the 
critical temperature gradient required for significant thermal transport, and can explain observed temperature gradients that significantly 
exceed the limits imposed by linear ETG turbulence.

Core NSTX plasmas with monotonic q-profiles exhibit ``stiff profiles": once the electron temperature gradient exceeds a critical value, 
turbulent heat flux relaxes the temperature profile back below the critical gradient \cite{wilson_exploration_2003}. In such cases, high-k
density fluctuation measurements show electron-scale turbulence \cite{mazzucato_short-scale_2008}. Additional electron 
heating cannot push the temperature gradient greatly beyond the linear threshold for ETG drive. Instead, the plasma profiles hover near 
marginal stability, with ETG turbulence limiting plasma performance.

However NSTX can violate the stiff profile phenomenon. Reversing the magnetic shear triggers an ``electron 
internal transport barrier'' (e-ITB) in NSTX \cite{yuh_internal_2009}. Interior to the e-ITB, electron temperatures can grow to very high 
values, while the temperature gradient at the e-ITB can reach values two to three times greater than the ETG linear critical gradient. 
Additionally, turbulent fluctuations and thermal transport drop \cite{mazzucato_study_2009, yuh_suppression_2011}. An NSTX plasma 
profile with an e-ITB is not stiff around the linear ETG critical gradient, but instead at much higher values.

The present study asks whether a nonlinearly enhanced critical gradient can explain 
NSTX reversed shear e-ITBs. A nonlinear upshift of the critical gradient exists for ion-temperature-gradient-driven (ITG) turbulence \cite{dimits_comparisons_2000, belli_effects_2008, 
mikkelsen_dimits_2008}, which can allow the normalized ion temperature gradient to increase by roughly $50\%$ prior to the onset of turbulence.  While early simulations of ETG turbulence 
showed low transport for $\hat{s}<0$
\cite{dorland_electron_2000, jenko_prediction_2002}, these studies found no enhanced critical gradient for the onset of turbulence;
turbulent transport remained low for all tested gradients. However, these works employed parameters not like those of NSTX.
Besides aspect ratio, NSTX can have much higher gradients and much lower values of $\hat{s}$ than those explored in Refs.~\cite{dorland_electron_2000, jenko_prediction_2002}.
In this context, we present the first nonlinear electron-scale gyrokinetic simulations of NSTX e-ITB plasmas, using 
multiple driving temperature gradients and values of magnetic shear. 

Herein, we find that reversed magnetic shear, without $\mathbf{E}\times\mathbf{B}$ shear, can nonlinearly reduce electron thermal transport up to a critical temperature gradient that is many times larger than the linear critical gradient, consistent with experimental observations \cite{yuh_internal_2009, yuh_suppression_2011}. We find the effect of reversed shear on ETG turbulence to be much more drastic than found that for ITG: the nonlinear critical gradient can reach three times the linear threshold, consistent with observations of NSTX e-ITBs.

Our simulations are based upon TRANSP \cite{budny_simulations_1992} analysis of NSTX shot number 129354 at 232 ms, during a radio-frequency-heated e-ITB phase. Experimental details are found in Refs.~\cite{yuh_internal_2009} and \cite{yuh_suppression_2011}. We use GYRO \cite{candy_gyro_2010} in a flux tube geometry \cite{beer_field-aligned_1995} with a Miller equilibrium \cite{miller_noncircular_1998} at the location of the e-ITB. Local parameters include:  $r/a=0.3,~R/a=1.55,~\Delta=-0.27,~\kappa=1.76,~s_\kappa=-0.35,~\delta=0.11,~s_\delta=-0.073,~\zeta=-2.0\times10^{-3},~s_\zeta=0.13,~q=2.40,~\hat{s}=-2.40,~\rho_\star=4.6\times10^{-2},~Z_{eff}=3.39$. While Ref.~\cite{candy_gyro_2010} contains full variable definitions, some key elements are: the device major radius, $R$; the full minor radius, $a$; the ion sound speed, $c_s=\sqrt{T_e/m_D}$; the ion sound radius, $\rho_s=c_s(m_Dc/eB_{unit})$; the electron thermal velocity, $v_e=\sqrt{T_e/m_e}$; the electron gyroradius, $\rho_e=v_e(m_ec/eB_{unit})$; the gyroBohm unit diffusivity, $\chi_{GB,i}=\rho_s^2c_s/a$; and the gyroBohm energy flux, $Q_{GB,i}=n_ec_sT_e(\rho_s/a)^2$. $n_{e/i}$, $T_{e/i}$, $m_{e/D}$, $c$ and $B_{unit}$ represent the background electron/ion density, temperature, the electron/deuterium mass, the speed of light and the effective magnetic field strength, respectively. Gradient lengths are given by the logarithmic derivative with respect to the minor radius: $L_T=(-T)/(\partial T/\partial r)$.

The primary ion is cold, low-density deuterium, treated gyrokinetically: $n_i/n_e=0.39,~T_i/T_e=0.22,~R/L_{n_i}=1.43,~R/L_{T_i}=2.65$. We use a realistic electron mass ratio, $\mu_e=\sqrt{m_i/m_e}=60$, and the electron density scale length remains fixed at $R/L_{n_e}=1.79$. The experimental value of the electron temperature gradient is $R/L_{T_e}=21.5\pm5$.
We retain electron-ion collisions, with $\nu_{ei}=0.016 (a/c_s)$, but neglect ion-ion collisions, background $\mathbf{E}\times
\mathbf{B}$ flow shear, 
and finite Debye length effects. Our 
simulations are all electrostatic.

We follow 24 toroidal modes with 256 radial grid points in a box with radial and binormal directions measuring $L_x\times 
L_y=4.26\times2.4\rho_s=255.6\times144\rho_e$, up to a maximum $k_\theta\rho_e$ of 1.004, $\left(k_\theta\rho_s
\right)_{max}=60.25$. Radial convergence required $\Delta x<\rho_e$. In agreement with linear eigenmode simulations of NSTX 
\cite{belli_fully_2010}, we needed increased resolution in other dimensions to reach nonlinear flux convergence. Details on parameters 
can be found in Refs.~\cite{candy_velocity-space_2006, candy_gyro_2010}, but briefly: $n_\tau=44$ (the number of mesh 
points along a trapped particle orbit), $n_b=12$ (the number of parallel finite elements), $n_\lambda=12$ (the number of pitch angles 
for each sign of parallel velocity, split evenly between trapped and passing particles), $n_\epsilon=12$ (the number of energy grid 
points) and $\epsilon^*=6.0$ (the maximum simulated dimensionless energy).
Temporal stability and convergence required time steps of $\Delta t\left(c_s/a\right)=3.3\times10^{-4}$, and simulation times of $t_{max}
\left(c_s/a\right)=10-50$. Each simulation costs 80,000-100,000 CPU hours on the ORNL Cray XT system, totaling in all over 5 million CPU hours.

To probe the existence of a nonlinearly enhanced threshold for the onset of electron transport, we scan $R/L_{T_e}$ from just above the linear critical gradient, of roughly $6$, well past physically-relevant values, to over $25$, for three different values of magnetic shear, which roughly cover NSTX's reversed shear operating space: $\hat{s}=-2.4$ (base parameter), $-1.0$ and $-0.2$.

Primarly, we find a nonlinear critical gradient for ETG thermal transport that is far larger than the linear critical gradient. 
For the baseline case of $\hat{s}=-2.4$, the linear critical gradient for turbulence is where the growth rate becomes positive, $z_{c}\doteq(R/L_{T_e})_{critical,~linear}\approx6$. Yet the heat flux remains low for driving gradients below $z_{c}^{NL}\doteq\left(R/L_{T_e}\right)_{critical,~nonlinear}\approx19$, the effective nonlinear critical gradient for transport. The difference, $\Delta z=z_{c}^{NL}-z_c$, represents the nonlinear upshift of the critical gradient for transport. In this case, $\Delta z\approx13$.

\begin{figure}
\includegraphics[]{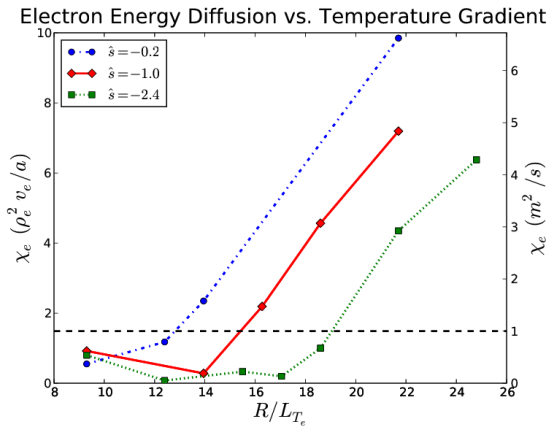}
\caption{(color online) Electron heat diffusivities for different driving gradients and three values of magnetic shear. The horizontal line at $\chi_e=1~m^2/s$ indicates experimentally relevant levels.}
\label{fig:chi_e_v_LTe_s}
\end{figure}

\begin{figure}
\includegraphics[width=0.9\columnwidth]{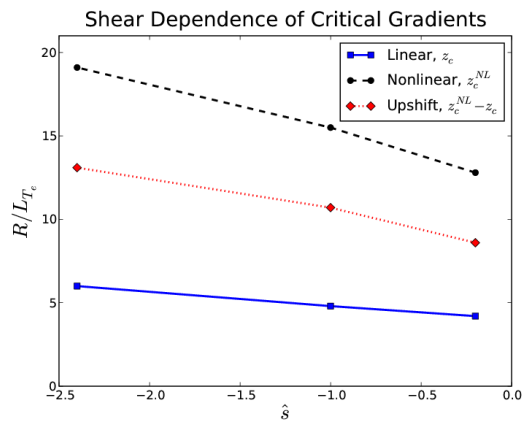}
\caption{(color online) Critical gradients as a function of magnetic shear. ETG becomes linearly unstable at gradients above $z_c$ (solid with square). Above $z_c^{NL}$ (dashed with circles), turbulent $\chi_e$ exceeds experimental values, $1~m^2/s$. The dotted line with diamonds represents the nonlinear upshift of the critical gradient, $\Delta z=z_c^{NL}-z_c$. For comparison, the original cyclone ITG test case found an upshift in the critical gradient for transport that extended $R/L_{T_i}$ by $2$, from $z_c=4$ to $z_c^{NL}=6$ at $\hat{s}=0.78$~\cite{dimits_comparisons_2000}. }
\label{fig:grads_v_shear}
\end{figure}

The strength of this upshift depends upon magnetic shear. Figure~\ref{fig:chi_e_v_LTe_s} shows electron heat diffusivities as a function of $R/L_{T_e}$ and $\hat{s}$. The horizontal dashed line corresponds to experimentally relevant levels, at $\chi_e=1~m^2/s$. We define $z_{c}^{NL}$ 
as the gradient above which $\chi_e$ exceeds this value, and plot this gradient and the corresponding values of $z_c$ for the cases in Fig.~\ref{fig:grads_v_shear}. As $\hat{s}$ decreases, $z_c$, $z_c^{NL}$ and $\Delta z$ increase, up to very large values: for the baseline value of $\hat{s}=-2.4$, $z_c^{NL}\approx19>3z_c$.

\begin{figure}
\includegraphics[]{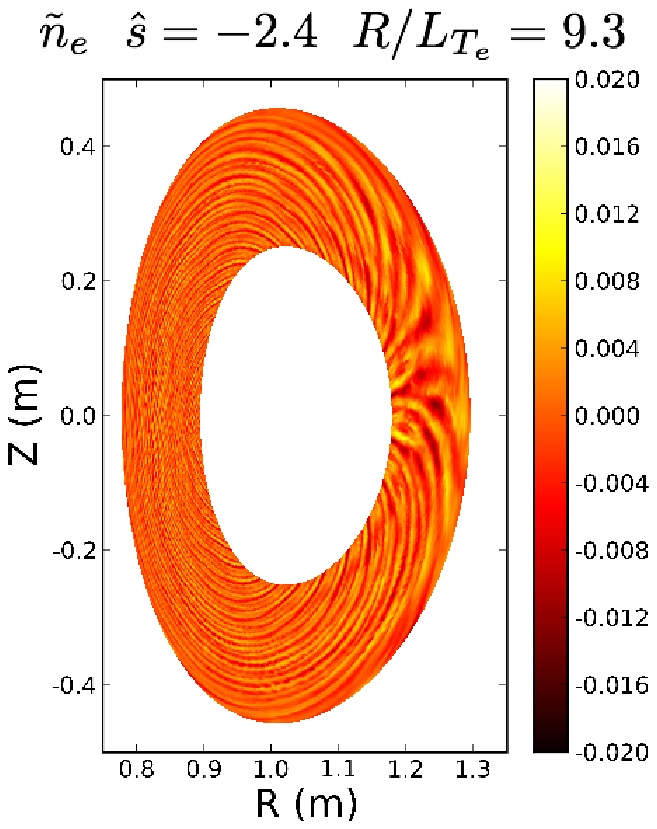}
\includegraphics[]{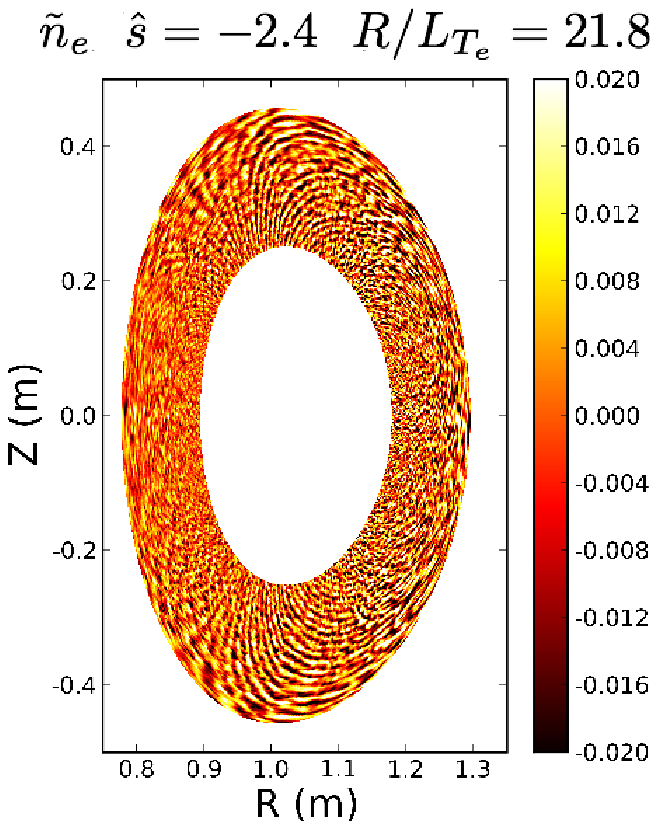}
\caption{(color online) Poloidal cross sections of saturated density fluctuations below and above $z_c^{NL}$ for $\hat{s}=-2.4$.}
\label{fig:LTE-contour-all}
\end{figure}
The nature of the saturated plasma turbulence changes drastically above and below $z_c^{NL}$. Figure~\ref{fig:LTE-contour-all} shows a poloidal cross section of computed electron density fluctuations above and below $z_c^{NL}$. For clarity we show only $\left|\tilde{n}_e\right|<2\%$. The top of Fig.~\ref{fig:LTE-contour-all} shows that below $z_c^{NL}$, fluctuations appear on the midplane. Their amplitude is small ($\left<\tilde{n}_e\right>_{rms}\approx0.3\%,~\left|\tilde{n}_e\right|_{max}\approx1.3\%$), and the sheared structure of the streamers is clearly visible. Eddies quickly rotate towards the vertical, thereby limiting their effect on transport, consistent with earlier investigations of reversed magnetic shear \cite{antonsen_physical_1996, jenko_prediction_2002}.
Above $z_c^{NL}$, a very different picture emerges, as in the top of Fig.~\ref{fig:LTE-contour-all}. Not only are these fluctuations stronger ($\left<\tilde{n}_e\right>_{rms}\approx1.1\%,~\left|\tilde{n}_e\right|_{max}\approx6.0\%$), their behavior is drastically different. Streamers appear out of the top and bottom of the flux tube, instead of along the midplane.

\begin{figure}
\includegraphics[]{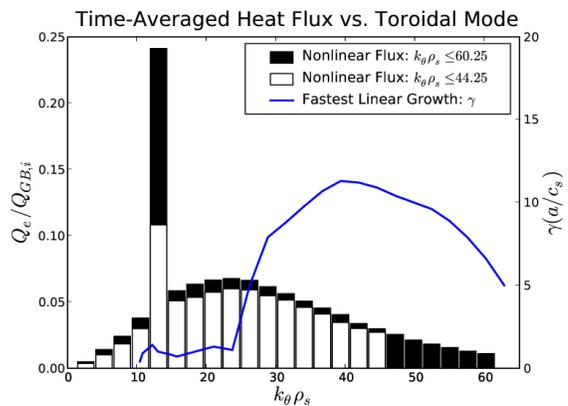}
\caption{(color online) Time-averaged heat flux spectra for two values of $(k_\theta\rho_s)_{max}$, at $R/L_{T_e}=21.8$, and the fastest growing linear mode.  $t_{avg}=[4.0, 8.0]~(a/c_s)$.}
\label{fig:Qe_n_boxes}
\end{figure}

\begin{figure}
\includegraphics[]{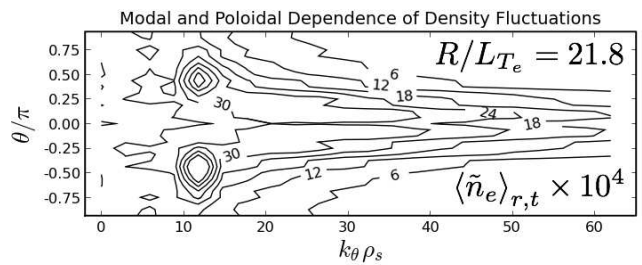}
\caption{Density fluctuations as a function of $\theta$ and $k_\theta\rho_s$, showing off-midplane peaking in $k_\theta\rho_s=13$.}
\label{fig:n_k_theta}
\end{figure}

The off-midplane streamers are localized to a narrow spectral band and appear to be nonlinearly driven by the peak of the linear growth 
spectrum. Figure~\ref{fig:Qe_n_boxes} shows the linear spectrum and time-averaged nonlinear heat flux per toroidal mode for two different values of $(k_\theta\rho_s)_{max}$. Though the fastest growing mode is centered near $k_\theta\rho_s=40$, the late-time heat flux consists of a broad spectrum about $k_\theta
\rho_s=23$ and a strong peak, near $k_\theta\rho_s=13$. The poloidal dependence of density fluctuations in Fig.~\ref{fig:n_k_theta} shows that, while the broad spectrum centers on the midplane, the spike is the heat flux from the off-midplane streamers. 
By reducing $k_{max}$, this heat flux drops, while the background spectrum is  modestly affected. This implies not only nonlinear energy cascade to the broad 
background turbulence, but also preferential coupling to the off-midplane streamers. We note that both the poloidal angle of the off-midplane streamers and their position in k-space vary with temperature gradient and shear.
Above the upshifted nonlinear critical gradient, we find that nearly one quarter of the turbulent transport is caused by a single slowly-growing mode driven to high amplitudes by the spectrum of faster growing, higher wavenumber modes.

The enhanced upshift of the critical gradient for ETG transport might be explained by secondary instability theory 
\cite{cowley_considerations_1991, dorland_electron_2000, jenko_prediction_2002, plunk_gyrokinetic_2007}. Primary instabilities grow in the plasma until they are balanced by a 
growth in a secondary mode, which serves as an energy sink for the primary mode. The growth of the secondary mode depends not only upon the properties of the 
primary mode (such as amplitude and wavenumber), but also upon some properties of the background plasma, such as $R/L_{T_e}$ and $\hat{s}$.
Gyrofluid estimates \cite{dorland_electron_2000} suggest negative magnetic shear reduces ETG transport by enhancing the damping due to the secondary mode. While this motivates the reduction in transport in these NSTX cases, to fully understand this strong upshift in the critical gradient, it would be useful to extend the work on 
gyrokinetic secondary theory \cite{plunk_gyrokinetic_2007} to reversed shear cases. An interesting note is that we, 
like Refs.~\cite{dorland_electron_2000, jenko_prediction_2002}, 
find low transport for adiabatic ion (ETG-ai) simulations. This is the opposite behavior than the ETG-ai transport runaway found at high positive shear \cite{candy_plasma_2007}. 
However, kinetic ions can affect the growth rate of the gyrokinetic ETG secondary instability \cite{plunk_gyrokinetic_2007}. 
The discovery of an ETG-ki nonlinear 
critical gradient further motivates  re-examining this theory in the context of reversed shear ETG turbulence.

In summary, our goal was to numerically investigate the presence of e-ITBs during reversed shear NSTX discharges. 
Our first-of-a-kind nonlinear ETG simulations of NSTX e-ITBs show that reversed magnetic shear, even in the 
absence of background $\textbf{E}\times\textbf{B}$ flow shear, is very effective at suppressing electron transport by significantly enhancing the nonlinear critical gradient for ETG turbulent flux. The upshifted gradient, which can approach 
three times the linear critical gradient, is much larger than predicted upshifts for ITG turbulence \cite{dimits_comparisons_2000, belli_effects_2008, mikkelsen_dimits_2008}.

Above this upshifted threshold, the turbulence takes on a different form than that first associated with ETG turbulence \cite{dorland_electron_2000, jenko_prediction_2002}. Characterized instead by off-midplane streamers, this variation in structure could have implications for high-k 
density scattering measurements during these discharges. Interpreting these results with a synthetic high-k diagnostic (such as in Ref.~\cite{poli_synthetic_2010}) 
and experimentally looking for these structures (by moving the high-k measurement location, for instance), 
could help elucidate the ``bursty" turbulence measured in some reversed shear NSTX e-ITBs \cite{yuh_suppression_2011}.

These results help explain not only NSTX e-ITBs, but also observations on other magnetic fusion devices. The RFX-Mod Reversed Field Pinch has found e-ITBs with non-monotonic q
profiles \cite{gobbin_vanishing_2011}, a feature also linked to ITB formation in tokamaks \cite{levinton_improved_1995,
strait_enhanced_1995, mazzucato_turbulent_1996, fujita_internal_1997, eriksson_discharges_2002}. As stellarators' external coils 
dictate the magnetic topology, it may be possible to optimize stellarator designs for e-ITBs. In short, shear reversal, even without 
background flow shear, could allow magnetic fusion devices to sustain electron temperature gradients much larger than the limits imposed by the linear onset of instability.

We thank E. Mazzucato, D. Smith and Y. Ren for the high-k measurements that motivated this work, the support of the SciDAC Center for the Study of Plasma Microturbulence, DOE Contract DE-AC02-09CH11466, and
the National Center for Computational Sciences at ORNL, under DOE Contract DE-AC05-00OR22725.


\begin{thebibliography}{30}
\expandafter\ifx\csname natexlab\endcsname\relax\def\natexlab#1{#1}\fi
\expandafter\ifx\csname bibnamefont\endcsname\relax
  \def\bibnamefont#1{#1}\fi
\expandafter\ifx\csname bibfnamefont\endcsname\relax
  \def\bibfnamefont#1{#1}\fi
\expandafter\ifx\csname citenamefont\endcsname\relax
  \def\citenamefont#1{#1}\fi
\expandafter\ifx\csname url\endcsname\relax
  \def\url#1{\texttt{#1}}\fi
\expandafter\ifx\csname urlprefix\endcsname\relax\def\urlprefix{URL }\fi
\providecommand{\bibinfo}[2]{#2}
\providecommand{\eprint}[2][]{\url{#2}}

\bibitem[{\citenamefont{Ono et~al.}(2000)}]{ono_exploration_2000}
\bibinfo{author}{\bibfnamefont{M.}~\bibnamefont{Ono}} \bibnamefont{et~al.},
  \bibinfo{journal}{Nucl. Fusion} \textbf{\bibinfo{volume}{40}},
  \bibinfo{pages}{557} (\bibinfo{year}{2000}).

\bibitem[{\citenamefont{Yuh et~al.}(2011)}]{yuh_suppression_2011}
\bibinfo{author}{\bibfnamefont{H.~Y.} \bibnamefont{Yuh}} \bibnamefont{et~al.},
  \bibinfo{journal}{Phys. Rev. Lett.} \textbf{\bibinfo{volume}{106}},
  \bibinfo{pages}{055003} (\bibinfo{year}{2011}).

\bibitem[{\citenamefont{Kaye
  et~al.}(2007{\natexlab{a}})}]{kaye_confinement_2007}
\bibinfo{author}{\bibfnamefont{S.}~\bibnamefont{Kaye}} \bibnamefont{et~al.},
  \bibinfo{journal}{Nucl. Fusion} \textbf{\bibinfo{volume}{47}},
  \bibinfo{pages}{499} (\bibinfo{year}{2007}{\natexlab{a}}).

\bibitem[{\citenamefont{Kaye et~al.}(2007{\natexlab{b}})}]{kaye_scaling_2007}
\bibinfo{author}{\bibfnamefont{S.~M.} \bibnamefont{Kaye}} \bibnamefont{et~al.},
  \bibinfo{journal}{Phys. Rev. Lett.} \textbf{\bibinfo{volume}{98}},
  \bibinfo{pages}{175002} (\bibinfo{year}{2007}{\natexlab{b}}).

\bibitem[{\citenamefont{Yuh et~al.}(2009)}]{yuh_internal_2009}
\bibinfo{author}{\bibfnamefont{H.~Y.} \bibnamefont{Yuh}} \bibnamefont{et~al.},
  \bibinfo{journal}{Phys. Plasmas} \textbf{\bibinfo{volume}{16}},
  \bibinfo{pages}{056120} (\bibinfo{year}{2009}).

\bibitem[{\citenamefont{Wilson et~al.}(2003)}]{wilson_exploration_2003}
\bibinfo{author}{\bibfnamefont{J.~R.} \bibnamefont{Wilson}}
  \bibnamefont{et~al.}, \bibinfo{journal}{Phys. Plasmas}
  \textbf{\bibinfo{volume}{10}}, \bibinfo{pages}{1733} (\bibinfo{year}{2003}).

\bibitem[{\citenamefont{Mazzucato et~al.}(2008)}]{mazzucato_short-scale_2008}
\bibinfo{author}{\bibfnamefont{E.}~\bibnamefont{Mazzucato}}
  \bibnamefont{et~al.}, \bibinfo{journal}{Phys. Rev. Lett.}
  \textbf{\bibinfo{volume}{101}}, \bibinfo{pages}{075001}
  (\bibinfo{year}{2008}).

\bibitem[{\citenamefont{Mazzucato et~al.}(2009)}]{mazzucato_study_2009}
\bibinfo{author}{\bibfnamefont{E.}~\bibnamefont{Mazzucato}}
  \bibnamefont{et~al.}, \bibinfo{journal}{Nucl. Fusion}
  \textbf{\bibinfo{volume}{49}}, \bibinfo{pages}{055001}
  (\bibinfo{year}{2009}).

\bibitem[{\citenamefont{Dimits et~al.}(2000)}]{dimits_comparisons_2000}
\bibinfo{author}{\bibfnamefont{A.~M.} \bibnamefont{Dimits}}
  \bibnamefont{et~al.}, \bibinfo{journal}{Phys. Plasmas}
  \textbf{\bibinfo{volume}{7}}, \bibinfo{pages}{969} (\bibinfo{year}{2000}).

\bibitem[{\citenamefont{Belli et~al.}(2008)\citenamefont{Belli, Hammett, and
  Dorland}}]{belli_effects_2008}
\bibinfo{author}{\bibfnamefont{E.~A.} \bibnamefont{Belli}},
  \bibinfo{author}{\bibfnamefont{G.~W.} \bibnamefont{Hammett}},
  \bibnamefont{and} \bibinfo{author}{\bibfnamefont{W.}~\bibnamefont{Dorland}},
  \bibinfo{journal}{Phys. Plasmas} \textbf{\bibinfo{volume}{15}},
  \bibinfo{pages}{092303} (\bibinfo{year}{2008}).

\bibitem[{\citenamefont{Mikkelsen and Dorland}(2008)}]{mikkelsen_dimits_2008}
\bibinfo{author}{\bibfnamefont{D.~R.} \bibnamefont{Mikkelsen}}
  \bibnamefont{and} \bibinfo{author}{\bibfnamefont{W.}~\bibnamefont{Dorland}},
  \bibinfo{journal}{Phys. Rev. Lett.} \textbf{\bibinfo{volume}{101}},
  \bibinfo{pages}{135003} (\bibinfo{year}{2008}).

\bibitem[{\citenamefont{Dorland et~al.}(2000)}]{dorland_electron_2000}
\bibinfo{author}{\bibfnamefont{W.}~\bibnamefont{Dorland}} \bibnamefont{et~al.},
  \bibinfo{journal}{Phys. Rev. Lett.} \textbf{\bibinfo{volume}{85}},
  \bibinfo{pages}{5579} (\bibinfo{year}{2000}).

\bibitem[{\citenamefont{Jenko and Dorland}(2002)}]{jenko_prediction_2002}
\bibinfo{author}{\bibfnamefont{F.}~\bibnamefont{Jenko}} \bibnamefont{and}
  \bibinfo{author}{\bibfnamefont{W.}~\bibnamefont{Dorland}},
  \bibinfo{journal}{Phys. Rev. Lett.} \textbf{\bibinfo{volume}{89}},
  \bibinfo{pages}{225001} (\bibinfo{year}{2002}).

\bibitem[{\citenamefont{Budny et~al.}(1992)}]{budny_simulations_1992}
\bibinfo{author}{\bibfnamefont{R.}~\bibnamefont{Budny}} \bibnamefont{et~al.},
  \bibinfo{journal}{Nucl. Fusion} \textbf{\bibinfo{volume}{32}},
  \bibinfo{pages}{429} (\bibinfo{year}{1992}).

\bibitem[{\citenamefont{Candy and Belli}(2010)}]{candy_gyro_2010}
\bibinfo{author}{\bibfnamefont{J.}~\bibnamefont{Candy}} \bibnamefont{and}
  \bibinfo{author}{\bibfnamefont{E.~A.} \bibnamefont{Belli}},
  \bibinfo{type}{General Atomics Report} \bibinfo{number}{{GA-A26818}},
  \bibinfo{institution}{General Atomics} (\bibinfo{year}{2010}).

\bibitem[{\citenamefont{Beer et~al.}(1995)\citenamefont{Beer, Cowley, and
  Hammett}}]{beer_field-aligned_1995}
\bibinfo{author}{\bibfnamefont{M.~A.} \bibnamefont{Beer}},
  \bibinfo{author}{\bibfnamefont{S.~C.} \bibnamefont{Cowley}},
  \bibnamefont{and} \bibinfo{author}{\bibfnamefont{G.~W.}
  \bibnamefont{Hammett}}, \bibinfo{journal}{Phys. Plasmas}
  \textbf{\bibinfo{volume}{2}}, \bibinfo{pages}{2687} (\bibinfo{year}{1995}).

\bibitem[{\citenamefont{Miller et~al.}(1998)}]{miller_noncircular_1998}
\bibinfo{author}{\bibfnamefont{R.~L.} \bibnamefont{Miller}}
  \bibnamefont{et~al.}, \bibinfo{journal}{Phys. Plasmas}
  \textbf{\bibinfo{volume}{5}}, \bibinfo{pages}{973} (\bibinfo{year}{1998}).

\bibitem[{\citenamefont{Belli and Candy}(2010)}]{belli_fully_2010}
\bibinfo{author}{\bibfnamefont{E.~A.} \bibnamefont{Belli}} \bibnamefont{and}
  \bibinfo{author}{\bibfnamefont{J.}~\bibnamefont{Candy}},
  \bibinfo{journal}{Phys. Plasmas} \textbf{\bibinfo{volume}{17}},
  \bibinfo{pages}{112314} (\bibinfo{year}{2010}).

\bibitem[{\citenamefont{Candy and Waltz}(2006)}]{candy_velocity-space_2006}
\bibinfo{author}{\bibfnamefont{J.}~\bibnamefont{Candy}} \bibnamefont{and}
  \bibinfo{author}{\bibfnamefont{R.~E.} \bibnamefont{Waltz}},
  \bibinfo{journal}{Phys. Plasmas} \textbf{\bibinfo{volume}{13}},
  \bibinfo{pages}{032310} (\bibinfo{year}{2006}).

\bibitem[{\citenamefont{Antonsen et~al.}(1996)}]{antonsen_physical_1996}
\bibinfo{author}{\bibfnamefont{T.~M.} \bibnamefont{Antonsen}}
  \bibnamefont{et~al.}, \bibinfo{journal}{Phys. Plasmas}
  \textbf{\bibinfo{volume}{3}}, \bibinfo{pages}{2221} (\bibinfo{year}{1996}).

\bibitem[{\citenamefont{Cowley et~al.}(1991)\citenamefont{Cowley, Kulsrud, and
  Sudan}}]{cowley_considerations_1991}
\bibinfo{author}{\bibfnamefont{S.~C.} \bibnamefont{Cowley}},
  \bibinfo{author}{\bibfnamefont{R.~M.} \bibnamefont{Kulsrud}},
  \bibnamefont{and} \bibinfo{author}{\bibfnamefont{R.}~\bibnamefont{Sudan}},
  \bibinfo{journal}{Phys. Fluids B: Plasma Phys.} \textbf{\bibinfo{volume}{3}},
  \bibinfo{pages}{2767} (\bibinfo{year}{1991}).

\bibitem[{\citenamefont{Plunk}(2007)}]{plunk_gyrokinetic_2007}
\bibinfo{author}{\bibfnamefont{G.}~\bibnamefont{Plunk}},
  \bibinfo{journal}{Phys. Plasmas} \textbf{\bibinfo{volume}{14}},
  \bibinfo{pages}{112308} (\bibinfo{year}{2007}).

\bibitem[{\citenamefont{Candy et~al.}(2007)}]{candy_plasma_2007}
\bibinfo{author}{\bibfnamefont{J.}~\bibnamefont{Candy}} \bibnamefont{et~al.},
  \bibinfo{journal}{J. Phys. Conf. Ser.} \textbf{\bibinfo{volume}{78}},
  \bibinfo{pages}{012008} (\bibinfo{year}{2007}).

\bibitem[{\citenamefont{Poli et~al.}(2010)}]{poli_synthetic_2010}
\bibinfo{author}{\bibfnamefont{F.~M.} \bibnamefont{Poli}} \bibnamefont{et~al.},
  \bibinfo{journal}{Phys. Plasmas} \textbf{\bibinfo{volume}{17}},
  \bibinfo{pages}{112514} (\bibinfo{year}{2010}).

\bibitem[{\citenamefont{Gobbin et~al.}(2011)}]{gobbin_vanishing_2011}
\bibinfo{author}{\bibfnamefont{M.}~\bibnamefont{Gobbin}} \bibnamefont{et~al.},
  \bibinfo{journal}{Phys. Rev. Lett.} \textbf{\bibinfo{volume}{106}},
  \bibinfo{pages}{025001} (\bibinfo{year}{2011}).

\bibitem[{\citenamefont{Levinton et~al.}(1995)}]{levinton_improved_1995}
\bibinfo{author}{\bibfnamefont{F.~M.} \bibnamefont{Levinton}}
  \bibnamefont{et~al.}, \bibinfo{journal}{Phys. Rev. Lett.}
  \textbf{\bibinfo{volume}{75}}, \bibinfo{pages}{4417} (\bibinfo{year}{1995}).

\bibitem[{\citenamefont{Strait et~al.}(1995)}]{strait_enhanced_1995}
\bibinfo{author}{\bibfnamefont{E.~J.} \bibnamefont{Strait}}
  \bibnamefont{et~al.}, \bibinfo{journal}{Phys. Rev. Lett.}
  \textbf{\bibinfo{volume}{75}}, \bibinfo{pages}{4421} (\bibinfo{year}{1995}).

\bibitem[{\citenamefont{{E. Mazzucato}
  et~al.}(1996)}]{mazzucato_turbulent_1996}
\bibinfo{author}{\bibnamefont{{E. Mazzucato}}} \bibnamefont{et~al.},
  \bibinfo{journal}{Phys. Rev. Lett.} \textbf{\bibinfo{volume}{77}},
  \bibinfo{pages}{3145} (\bibinfo{year}{1996}).

\bibitem[{\citenamefont{Fujita et~al.}(1997)}]{fujita_internal_1997}
\bibinfo{author}{\bibfnamefont{T.}~\bibnamefont{Fujita}} \bibnamefont{et~al.},
  \bibinfo{journal}{Phys. Rev. Lett.} \textbf{\bibinfo{volume}{78}},
  \bibinfo{pages}{2377} (\bibinfo{year}{1997}).

\bibitem[{\citenamefont{Eriksson et~al.}(2002)}]{eriksson_discharges_2002}
\bibinfo{author}{\bibfnamefont{L.}~\bibnamefont{Eriksson}}
  \bibnamefont{et~al.}, \bibinfo{journal}{Phys. Rev. Lett.}
  \textbf{\bibinfo{volume}{88}}, \bibinfo{pages}{145001}
  (\bibinfo{year}{2002}).

\end{thebibliography}


\end{document}